# UNSUPERVISED EMPIRICAL BAYESIAN MULTIPLE TESTING WITH EXTERNAL COVARIATES

By Egil Ferkingstad,[1] Arnoldo Frigessi, Håvard Rue, Gudmar Thorleifsson and Augustine Kong

*University of Oslo and Centre for Integrative Genetics, $(sfi)^2$—Statistics for Innovation, Norwegian University of Science and Technology, Decode Genetics and Decode Genetics*

In an empirical Bayesian setting, we provide a new multiple testing method, useful when an additional covariate is available, that influences the probability of each null hypothesis being true. We measure the posterior significance of each test conditionally on the covariate and the data, leading to greater power. Using covariate-based prior information in an unsupervised fashion, we produce a list of significant hypotheses which differs in length and order from the list obtained by methods not taking covariate-information into account. Covariate-modulated posterior probabilities of each null hypothesis are estimated using a fast approximate algorithm. The new method is applied to expression quantitative trait loci (eQTL) data.

**1. Introduction.** Science, industry and business possess the technology to collect, store and distribute huge amounts of data efficiently and often at low cost. Sensors and instrumentation, data logging capacity and communication power have increased the breadth and depth of data. Systems are measured more in detail, giving a more complete but complex picture of processes and phenomena. Also, it is necessary to integrate many sources of data of different type and quality. In high-throughput genomics, large numbers of simultaneous comparisons are necessary to discover differentially expressed genes among thirty thousand measured ones. Similarly, in finance, one wishes to monitor prices of thousands of products and derivatives simultaneously to detect abnormal behavior, or in geophysics or brain imaging, questioning thousands of 3D voxels about their properties. Such tests are

Received March 2007; revised January 2008.
[1]Supported by the National program for research in functional genomics in Norway from the Research Council of Norway.

often dependent, and the dependency structure is ill specified, so that the effective number of independent tests is unknown. Sometimes, we expect that only a small subset of decisions will have a positive result: the solution is then sparse in the huge parameter space. To discover significant cases, it is necessary to develop new methods that either exploit available a priori knowledge on the structure of the solution, or merge different data sets, each adding information. Benjamini and Hochberg (1995) proposed the false discovery rate (FDR), which can adapt automatically to sparsity and has been shown to be asymptotically optimal in a certain minimax sense [Abramovich et al. (2006)]. FDR adjustments of p-values are nowadays routinely performed on large scale multiple studies in many sciences and applied areas, from astronomy [Miller et al. (2001)] to genomics [Tusher et al. (2001); from neuroimaging [Genovese, Lazar and Nichols (2002)] to industrial organization [Brown et al. (2005)]. Bayesian approaches are based on the estimation of the posterior probability of the null hypothesis. Efron et al. (2001) have developed the theory of the *local false discovery rate*, based on an estimation procedure originally developed by Anderson and Blair (1982). As the FDR provides a probability of misclassification for sets of tests called significant, the posterior probability that the null hypothesis is true provides a similar measure, but for a local set about the particular value of the test statistic. Instead of summarizing the data by a test statistic, hierarchical Bayesian approaches have been developed that model parametrically the full measured data [Baldi and Long (2001), Do, Müller and Tang (2005), Kendziorski et al. (2006), Lonnstedt and Speed (2002), Newton et al. (2004), and Storey (2007) also makes full use of the data in a hypothesis testing setting. Both approaches have their strengths and weaknesses, in terms of validity of the distributional assumptions under the alternative hypothesis, actual availability of the full data, computational speed and simplicity of the methodology. This paper assumes access to summary test statistics for every hypothesis to be tested.

We propose a simple methodology which allows modulating the posterior probability of each null hypothesis based on a priori additional external information. Assume that for each null hypothesis $H_{0i}$ a known covariate $x_i$ is available which could influence the prior probability that $H_{0i}$ is true. Typically, the covariates $x_i$ represent available knowledge deriving from other data and measurement technologies. Our method is unsupervised and semiparametric, in the sense that it is not necessary to model the joint distribution of the test statistics with the covariate. By means of empirical Bayes, we take advantage of prior information on the probability of each null hypothesis being true, based on such additional data available for each single test, to produce a more precise list of rejected null hypotheses. This leads to a measure of the posterior significance of each test $i$, conditional on the covariate $x_i$ and the data, often leading to greater power. Furthermore, the ranking of the hypotheses based on the covariate modulated posterior probability is different from the order provided in absence of the external covariate.

We estimate the covariate modulated posterior probability of each hypothesis by binning the data according to the external covariate, with the help of an approximate mixture model on p-values (Section 2). Inference is then based on approximate Bayesian inference [Rue and Martino (2007)], in order to reduce computational time (Section 3). In Section 4 we present the analysis of expression quantitative trait loci (eQTL) data, where gene expression measurements obtained by microarrays are combined with genetic linkage analysis. Our example explicitly considers an important case where (1) the test is not a comparison, (2) the covariate cannot be used for testing alone, and (3) there is no way (that we know of) to produce a full model. We then compare our approach with the recently proposed Optimal Discovery Procedure (ODP) [Storey (2007)], and show that our method is optimal in the sense of Storey, when data are available only at test statistic level, within each covariate-determined bin (Section 5). Section 6 includes a simulation study, which illustrates the effect of external information on the power of the test. Section 7 presents some discussion and extensions. Our method provides a simple unsupervised way of incorporating different information into the Bayesian hypothesis testing framework, when full modeling of the complete data is not practical or reliable.

**2. Covariate-modulated multiple testing.** We consider the simultaneous testing of $m$ hypotheses:

$$H_{0i} \text{ vs. } H_{1i}, \qquad i=1,\ldots,m.$$

For each $i$, a test statistic $Z_i = z_i$ is calculated. If $H_{0i}$ is true, then $Z_i$ has a known distribution $F_0$ [e.g., $F_0$ could be $N(0,1)$ or a t-distribution]. Here, large $Z_i$ are evidence against the null hypothesis. Also, for each $i$, we have a covariate $x_i$ which may influence the prior probability that the null hypothesis is true. We consider the $x_i$, $i=1,\ldots,m$, to be known covariates. We can reasonably assume that $x_i$ does not influence the probability distributions of $Z_i$ under $H_{0i}$, so $Z_i \sim F_0$ under $H_{0i}$ no matter what the value of $x_i$ is. However, for each $i$, the distribution of $Z_i$ under $H_{1i}$ depends on $x_i$. Assume that, for each $i$, $H_{1i} = H_{0i}^c$ in some space of interest.

Let $g(z_i|x_i)$ be the density of $Z_i$ given $x_i$. From the assumptions above, the following basic model is derived:

(1) $$g(z_i|x_i) = \pi_0(x_i)g(z_i|H_{0i}) + (1-\pi_0(x_i))g(z_i|H_{1i},x_i),$$

where $\pi_0(x_i) = P(H_{0i}|x_i)$. Here, $\pi_0(x_i)$ and $g(z_i|H_{1i},x_i)$, the density of $Z_i$ given $x_i$ under $H_{1i}$, are unknown, while the distribution function corresponding to $g(z_i|H_{0i})$ is known to be $F_0$.

For each test statistic $Z_i$, the corresponding p-value $P_i$ is given by $P_i = 1 - F_0(Z_i)$. Hence, based on the observed $z_i$, $i = 1, \ldots, m$, we can calculate observed p-values as

$$p_i = 1 - F_0(z_i), \qquad i = 1, \ldots, m.$$

We can write the basic model (1) in terms of p-values instead of z-scores. Let $f(p_i|x_i)$ be the density of p-value $p_i$ given covariate $x_i$. Clearly, the distribution of the p-value $P_i$ under $H_{0i}$ is uniform on $[0,1]$ and does not depend on $x_i$. Accordingly, the model in terms of p-values is

$$(2) \qquad f(p_i|x_i) = \pi_0(x_i) + (1 - \pi_0(x_i))f(p_i|H_{1i}, x_i).$$

We define the covariate-modulated posterior probability of $H_{0i}$ as $P(H_{0i}|p_i, x_i)$.

We have data $(p_i, x_i)$ and need to estimate $\pi_0(x_i)$ and $f(p_i|H_{1i}, x_i)$. Diaconis and Ylvisaker [(1985), Theorem 1] showed that any distribution on $[0,1]$ can be well approximated by a mixture of beta distributions. Allison et al. (2002) investigated this approach further and applied it to estimating the density $f$ underlying a sample of p-values. In their experience with several sets of data, the simplest possible model, which is a mixture of a standard uniform, $U[0,1]$, corresponding to the true null hypotheses and one beta component corresponding to the alternative hypotheses, seemed to be sufficient; furthermore, Parker and Rothenberg [(1988), Section 3] noted that adding more beta components will increase the number of observed rejections. Therefore, using only one single beta component corresponding to the false null hypotheses can be seen as a *conservative* choice, as we avoid overestimating the proportion of false null hypotheses. We have chosen to use a mixture of a uniform density and a beta density as our model.

To account for the dependence on $x_i$ in $f(p_i|x_i)$, we bin the p-values into $B$ sets $\mathcal{B}_1, \mathcal{B}_2, \ldots, \mathcal{B}_B$ increasing in $x$. We assume that bins are small enough so that $\pi_0(x)$ is nearly constant in $x$ for each bin $\mathcal{B}_j$, that is, $\pi_0(x) = \pi_{0j}$ for all $x \in \mathcal{B}_j$, and similarly for the parameters of the beta distribution. Thus, we drop the dependence on $x$ *within* each bin. We assume that within each bin $\mathcal{B}_j$, the uniform-beta mixture model holds, so

$$(3) \qquad f_j(p_i) = \pi_{0j} + (1 - \pi_{0j})\frac{\Gamma(\xi_j + \theta_j)}{\Gamma(\xi_j)\Gamma(\theta_j)}p_i^{\xi_j - 1}(1 - p_i)^{\theta_j - 1}$$

for $p_i \in \mathcal{B}_j$, where $\Gamma$ denotes the gamma function, and $\xi_j > 0$ and $\theta_j > 0$ are the parameters of the beta density in bin $\mathcal{B}_j$. The covariate-modulated posterior probability of $H_{0i}$ corresponding to a p-value $p_i$ in bin $\mathcal{B}_j$ is then

$$P(H_{0i}|p_i, x_i) = \frac{P(H_{0i}|x_i)}{f(p_i|x_i)} = \frac{\pi_{0j}}{f_j(p_i)}.$$

Since small (large) p-values should correspond to false (true) null hypotheses, p-value densities should be nonincreasing. Hence, we assume that, in each bin $j$, $f_j(p)$ is a nonincreasing function of $p$ for $p \in [0, 1]$. [See also Wu, Guan and Zhao (2006), who show that $f_j(p)$ is always decreasing for p-values from a likelihood ratio test.] In addition, we make the assumption that $f_j(p)$ is convex. This is done to avoid underestimation of $f_j(p)$ near $p = 1$, which would lead to underestimation of $\pi_{0j}$ [since $\pi_{0j} = f_j(1)$]. Without the convexity assumption, the assumption of non-increasingness for a density with bounded support may lead to underestimation due to a "drop-down effect" near the right endpoint. See Langaas, Lindqvist and Ferkingstad (2005) for further discussion of the convexity assumption. It can be shown that $f_j(p)$ is nonincreasing and convex if and only if $\xi_j \leq 1$ and $\theta_j \geq 2$.

At this point, we have a separate model for each bin. The next step is to smooth over the different bins and borrow strength between neighboring bins, leading to improved estimates. This can be done using a Bayesian approach, as follows.

Defining a smoothness prior on the sequence of $\pi_{01}, \ldots, \pi_{0B}$ is complicated by the fact that each $\pi_{0j}$ is limited to the interval $[0, 1]$. We address this issue using reparametrization

$$\widetilde{\pi}_{0j} = \log \frac{\pi_{0j}}{1 - \pi_{0j}},$$

so that $\widetilde{\pi}_{0j}$ is defined on the whole real line. The smoothness prior is now taken to be

$$(4) \qquad f(\widetilde{\pi}_{01}, \ldots, \widetilde{\pi}_{0B}) \propto \exp\left(-\frac{\lambda_1}{2} \sum_{j=2}^{B} (\widetilde{\pi}_{0j} - \widetilde{\pi}_{0(j-1)})^2\right)$$

to encourage the parameter values in neighboring bins to be similar. Here, $\lambda_1$ is the smoothing parameter. Note that (4) is improper as it is invariant to adding any constant to its arguments.

The remaining two parameter sequences, $\xi_1, \ldots, \xi_B$, and $\theta_1, \ldots, \theta_B$ have similar restrictions, as $\xi_j \in [0, 1]$ and $\theta_j > 2$, for each $j$. The smoothness priors for $\{\xi_j\}$ and $\{\theta_j\}$ are defined similarly as for $\{\pi_{0j}\}$, using the reparametrization

$$\widetilde{\xi}_j = \log \frac{\xi_j}{1 - \xi_j} \quad \text{and} \quad \widetilde{\theta}_j = \log(\theta_j - 2).$$

Denote the smoothing parameters as $\lambda_2$ and $\lambda_3$, respectively.

Now, denoting the p-values in bin $j$ by $p_{j_1}, p_{j_2}, \ldots, p_{j_{m_j}}$, the simultaneous posterior of interest is then

$$f(\{\widetilde{\pi}_{0j}\}, \{\widetilde{\xi}_j\}, \{\widetilde{\lambda}_j\} \mid \lambda_1, \lambda_2, \lambda_3, \text{data})$$
$$\propto f(\{\widetilde{\pi}_{0j}\}) f(\{\widetilde{\xi}_j\}) f(\{\widetilde{\theta}_j\})$$

(5)
$$\times \prod_{h=1}^{m_j} \biggl[ \pi_{0j}(\widetilde{\pi}_{0j}) + (1 - \pi_{0j}(\widetilde{\pi}_{0j})) $$
$$\times \frac{\Gamma(\xi_j(\widetilde{\xi}_j) + \theta_j(\widetilde{\theta}_j))}{\Gamma(\xi_j(\widetilde{\xi}_j))\Gamma(\theta_j(\widetilde{\theta}_j))} p_{j_h}^{\xi_j(\widetilde{\xi}_j)-1}(1 - p_{j_h})^{\theta_j(\widetilde{\theta}_j)-1} \biggr].$$

We perform inference on the transformed parameters as this is an unconstrained parametrization, but can easily transform back to the original parameters.

For nearest-neighborhood improper priors, smoothing is regulated by the smoothing parameters $(\lambda_1, \lambda_2, \lambda_3)$. These need to be estimated or tuned in calibration experiments. Among the various approaches for estimating smoothing parameters, we mention cross-validation [Thompson et al. (1991)], estimates based on approximate models, and fully Bayesian inference [Kunsch (1994), Heikkinen and Penttinen (1999)]. We find preliminary estimates of the parameters $\widetilde{\pi}_{0j}, \widetilde{\xi}_j, \widetilde{\theta}_j;\ j = 1, \ldots, B$, without smoothing; then we fit a Gaussian model to each parameter and estimate $(\lambda_1, \lambda_2, \lambda_3)$ based on the estimated inverse variances. Thus,

$$\hat{\lambda}_1 = B \Big/ \sum_{j=2}^{B} (\hat{\widetilde{\pi}}_{0j} - \hat{\widetilde{\pi}}_{0(j-1)})^2,$$

and $\hat{\lambda}_2$ and $\hat{\lambda}_3$ are defined similarly. In addition, we consider scaling each smoothing parameter by a tuning parameter $c > 0$, leading to $c\hat{\lambda}_1, c\hat{\lambda}_2, c\hat{\lambda}_3$. The eQTL data analysis in Section 4 is performed for both $c = 1$ and $c = 5$ for comparison.

It is easy to estimate the beta-mixture model from equation (3) with only one bin, $\mathcal{B} = [0, 1]$, thus discarding the information in the covariate. We will refer to this as the "one-bin model." The estimated posterior probabilities from the one-bin model are useful for assessing the usefulness of the information contained in the covariate, by comparing the number of rejections given by the one-bin and $B$-bin models.

Finally, note that the ordering of tests by significance will often be different for the covariate-modulated posterior probability method than for methods not taking the covariate into account, such as Efron's local FDR and the one-bin model. If, for example (as in the data set considered in Section 4), the covariate-modulation function $\pi_0(x)$ is decreasing in $x$, then tests corresponding to a low value of $x$ may move down in the significance list, and tests corresponding to a high $x$ may move up in the list. This reordering of significance will be illustrated for the eQTL data set analyzed in Section 4.

We estimate the $\pi_{0j}$, $\xi_j$, $\theta_j$ and the covariate-modulated posterior probability of the $H_{0i}$ simultaneously using the approximate Bayesian method described in Section 3.

**3. Computational strategy.** This section discusses how to develop an efficient strategy for doing inference from the full posterior of interest (5). Although inference is possible using Markov chain Monte Carlo methods, we need a computationally faster approach for our problem. The amount of data in each bin, combined with the Gaussian smoothing, justify to approximate the joint posterior of $\{\widetilde{\pi}_{0j}\}, \{\widetilde{\xi}_j\}, \{\widetilde{\theta}_j\}$ as Gaussian. The posterior mean is the modal configuration, and the posterior covariance matrix is the inverse of the negative Hessian at the mode. Note that the Gaussian approximation is likely to be less accurate without reparametrization, due to the constraints of the parameters.

To compute the posterior mean and the covariance matrix, we need to optimize the log posterior, that is, the log of (5). We do this in two steps; in the first step we compute reasonable initial values for the parameters in each of the $B$ bins, while in the second step we start from these initial values and optimize the log posterior with respect to all the parameters to locate the mode:

1. The initial values for the parameters of interest are determined sequentially. We start with the first bin. We use only the data in the first bin to optimize the log likelihood with respect to $\widetilde{\pi}_{01}, \widetilde{\xi}_1, \widetilde{\theta}_1$. Then we go on to the second bin using the initial values found from the first bin to initialize the optimization in bin 2. This process continues until all the $B$ bins are processed.
2. The log posterior is then optimized with respect to all the parameters starting from the initial values found in the previous step.

Many numerical optimization schemes can be used in each of the two steps, and we use the classical Newton–Raphson algorithm. The main motivation is that both the gradient and the Hessian of the log posterior (both in the full model and for each bin separately) are relatively easy to compute. Further, the Hessian matrix of the log posterior will be sparse in the full model since the parameters in each bin are only linked to the previous and following bin. As a side effect, we can invert the negative Hessian used at the last iteration to obtain the covariance matrix.

As all the parameters are (approximately) jointly Gaussian, then so are the posterior marginals for each $\widetilde{\pi}_{0j}$, $\widetilde{\xi}_{0j}$ and $\widetilde{\theta}_{0j}$, and they can be found analytically.

The posterior density of $P(H_{0i}|p_i, x_i)$ for bin $j$ can be computed from the joint posterior for $\widetilde{\pi}_{0j}, \widetilde{\xi}_j, \widetilde{\theta}_j$. Such an approach is feasible but cumbersome using numerical integration. For this reason, we attack this problem using the delta-method: Expand $P(H_{0i}|p_i,x_i)$ around the posterior mean of $(\widetilde{\pi}_{0j},\widetilde{\xi}_j,\widetilde{\theta}_j)$, $(\widetilde{\pi}_{0j}^*,\widetilde{\xi}_j^*,\widetilde{\theta}_j^*)$, say, to obtain

$$P(H_{0i}|p_i,x_i) = a + b^T((\widetilde{\pi}_{0j},\widetilde{\xi}_j,\widetilde{\theta}_j) - (\widetilde{\pi}_{0j}^*,\widetilde{\xi}_j^*,\widetilde{\theta}_j^*)) + \cdots.$$

We now approximate the posterior of $P(H_{0i}|p_i,x_i)$ at bin $j$, by a Gaussian with mean $a$ and variance $b^T \Sigma_j b$, where $\Sigma_j$ is the posterior covariance of $(\widetilde{\pi}_{0j},\widetilde{\xi}_j,\widetilde{\theta}_j)$.

We have verified our Gaussian approximations using long runs of an MCMC algorithm without being able to detect any relevant differences.

**4. eQTL data.** Through the collection of phenotypic and genetic data for individuals in family clusters, linkage analysis is a standard method to map genetic variants that can influence a trait to specific regions of the genome [Ott (1999)]. Recently, there was the recognition that the magnitude of gene expression, measured using modern microarray technology for many genes simultaneously, varies among individuals and often has a genetic component [Jansen and Nap (2001), Schadt et al. (2003), Brystrykh et al. (2005)]. It hence can be treated as a quantitative genetic trait. The loci affecting gene expression are referred to as expression quantitative trait loci (eQTLs). While the same linkage methodology applies, there are specific characteristics that distinguish the study of eQTLs from the study of other more traditional traits. First, often the expressions of tens of thousands of genes are studied simultaneously. Second, a variant that affects the expression of a certain gene that is located in the immediate neighborhood of where the gene resides is called a cis-variant, whereas a variant that can affect the expression of the gene, directly or indirectly, but is located far away (e.g., on a different chromosome), is a trans-variant. In general, given the phenotypic and genetic data, for each trait, a linkage score can be calculated for every position on the genome (referred to as a genome-scan). Here we focus on the test of whether a cis-variant exists so that only the linkage score evaluated at the known location of the gene is used. The data involve the expressions of 22317 expressed sequence tags (ESTs) for 370 Icelandic individuals clustered into 85 families. The data structure is similar to that of Morley et al. (2004) and Monks et al. (2004), the difference being that we have a larger sample size and the gene expressions are measured in blood instead of lymphoblastoid cell lines.

Denote the gene expression (phenotype) data by $E$ and the genotype data by $G$. Using linkage analysis, the joint data $(E,G)$ is used to test the null hypothesis

$$H_0: \text{No cis-variant is affecting } E,$$

through the conditional distribution $P(G|E)$, for each EST (index dropped here for simplicity). In a linkage analysis, as opposed to an association analysis, we do not use the correlation between the genotype data $G$ and $E$ directly. Rather, the genotype data are used to track the segments of chromosomes that are shared by relatives identical by descent. Evidence of linkage of a phenotype to a genomic region shows up when relatives having similar phenotypes share a region by descent in excess of what is expected based on their known relationships, and relatives with phenotypes that are substantially different have a deficit of sharing. The test we used is particularly simple. Under $H_0$, $G$ is independent of $E$, or $P_0(G|E) = P(G)$, and the distribution of any test statistic has a known distribution under $H_0$ with no nuisance parameters. Specifically, what we used here is an allele-sharing score, an extension of that described in Kong and Cox (1997) and closely related to the method of Sham et al. (2002). Through exponential tilting which leads to a one-parameter alternative distribution for the allele-sharing score, a simple likelihood ratio test is used for each EST. The linkage score is calculated using the program Allegro [Gudbjartsson et al. (2000)]. Simulation shows that the p-values calculated using the chi-square distribution as the reference are very well calibrated, as expected.

While linkage analysis involves the study of the co-segregation of phenotypes and genetic material, the phenotype data alone can be used to estimate heritability, that is, the strength of the correlation of phenotypes among relatives can be used to estimate the fraction of the phenotype variance that is potentially accounted for by genetic variants. Denote the estimated heritability by $H(E)$, which we compute from $E$ using the program SOLAR [Almasy and Blangero (1998)] for each of the 22317 ESTs.

It might appear that we have artificially partitioned the overall information captured by the joint distribution $P(G, E)$ into two parts, one based on the conditional distribution $P(G|E)$, and one based on $P(E)$. However, in this case, as in many other real data problems, there is substantial asymmetry between the information captured by $P(G|E)$ and $P(E)$. As noted above, the test we used to directly test $H_0$ and compute p-values is simple and straightforward. It is very different for $H(E)$. Specifically, $H(E)$ can be significantly and substantially different from zero for at least four different reasons:

(A) There exists cis-variants that affect $E$.
(B) There exist trans-variants (variants that are located in regions of the genome that are away from the gene for which the expression is measured) that, directly or indirectly, affect $E$.
(C) $H(E)$ is capturing familial clustering/similarities that result from shared environment among relatives instead of genetic factors.

(D) Subjects and families are often not collected completely at random. Nonrandom ascertainment with respect to traits such as obesity, which are associated with some of the expression phenotypes, could lead to bias in the estimate of heritability.

Reason (A) above is the alternative for $H_0$ as specified above, but $E$ or $H(E)$ cannot be used to directly test $H_0$ against the truth of (A). Therefore, we are not able to use $H(E)$ directly to compute p-values. However, there are obvious reasons to believe that the probability for (A) being true is correlated with $H(E)$. Making as little assumptions as possible, our approach is to use $H(E)$ through a semi-parametric empirical-Bayes procedure to provide a prior distribution for $H_0$ and (A).

Covariate-modulated posterior probabilities were calculated as described in Sections 2 and 3. Twenty bins were used, with bin 1 containing the p-values with corresponding heritability estimated exactly equal to zero, and the other bins chosen to contain an approximately equal number of p-values. Smoothing scaling factors $c = 1$ and $c = 5$ were used. Resulting estimates in four of the bins, for $c = 5$, with 0.95 pointwise symmetric credibility intervals, are shown in Figure 1. Here we see that, for any given value of $p$, the covariate-modulated posterior probability of $H_0$ decreases for increasing bin index (increasing heritability). This effect seems to be quite strong.

The one-bin model was used to assess the effects of covariate-modulation. Figure 2 shows a comparison between the results of the covariate-modulated posterior probability method and the one-bin, no-covariate method. This plot shows a 2-dimensional histogram of the (covariate, test statistic) pair for each of the 22317 ESTs. A higher density of tests (ESTs) is indicated by a darker gray tone in the histogram. Assume that a test is called significant if the posterior probability of the null hypothesis is smaller than 0.05. The step-like solid curve is the significance threshold at level 0.05 for the 20-bin covariate-modulated posterior probability based method. The dashed line shows the significance threshold, also at level 0.05, for the one-bin, noncovariate based, posterior probability. A total of 818 tests were called significant by both methods, while 704 tests were called significant by the 20-bin method only. 53 tests (corresponding to EST's with low heritability) were called significant by the one-bin method, but discarded as such by the covariate-modulated 20-bin method. These results are for smoothing scale $c = 5$. Results for smoothing scale $c = 1$ are similar, with 817 tests called significant for both methods, 738 tests called significant for the 20-bin method only, and 54 tests called significant for the one-bin method only.

The impact of using heritability information is huge. Looking at Figure 2, it is evident that there is a large difference when using the covariate-modulated method compared to the no-covariate method. In Figure 2 the 53 tests in the region below the 20-bin threshold and above the one-bin threshold are considered by the covariate-based method as false discoveries, obtained erroneously by a no-covariate method. These points are in a region of lower heritability, while the 704 points below the one-bin significance line and above the 20-bin significance curve (which are "gained" by introducing covariate-modulation) are in the region with higher heritability. As expected, the tests with low heritability are effectively down-weighted by the 20-bin method, while tests with high heritability are up-weighted, in a nonsupervised fashion by the Bayesian rule. Thus, there are two potential gains of using our method: Higher overall power, and better focus on individual findings supported by additional information.

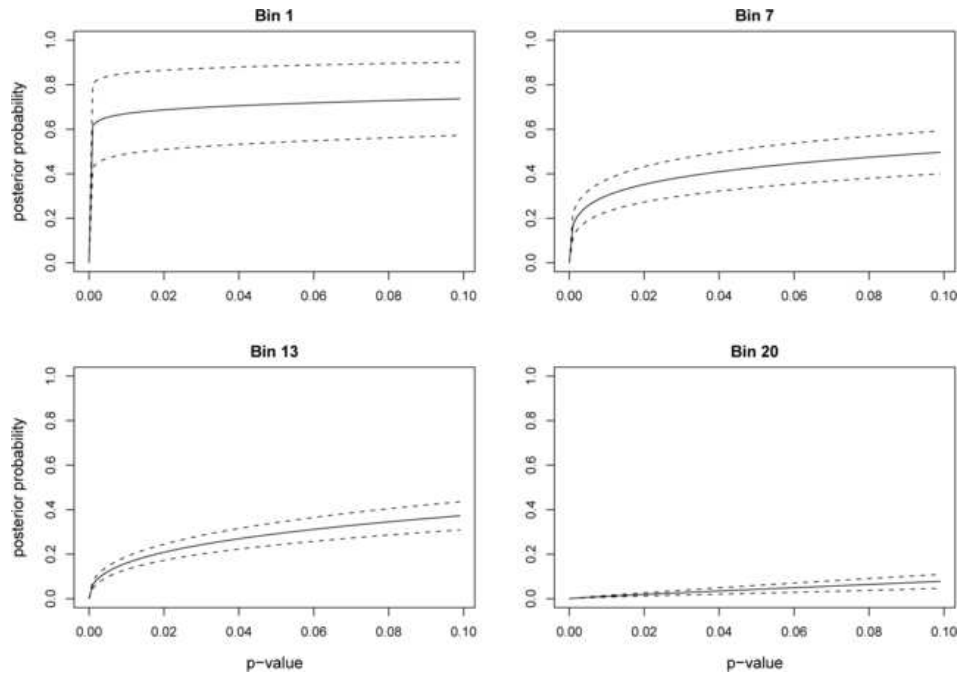

FIG. 1. *Estimated covariate-modulated posterior probabilities of $H_0$ in bins 1, 7, 13 and 20. The solid line shows the estimate, while the dashed lines are approximate pointwise 95% symmetric credibility intervals.*

It may also be of interest to study the estimated covariate-modulated functions (i.e., the estimated $\pi_{01}, \pi_{02}, \ldots, \pi_{0B}$) directly. A plot of $\pi_{0j}$ vs. covariate $x$ for is shown in Figure 3. Clearly, the dependence on the covariate, heritability, is very strong. The covariate-modulation function also illustrates the effect of smoothing. The above plot shows the case $c = 1$, while the below plot shows the case $c = 5$. In this case, smoothing $c = 5$ seems most appropriate, based on the degree of smoothness and the a priori expectation that the covariate-modulation function should be decreasing in $x$.

## 20-bin vs. 1-bin covariate modulation

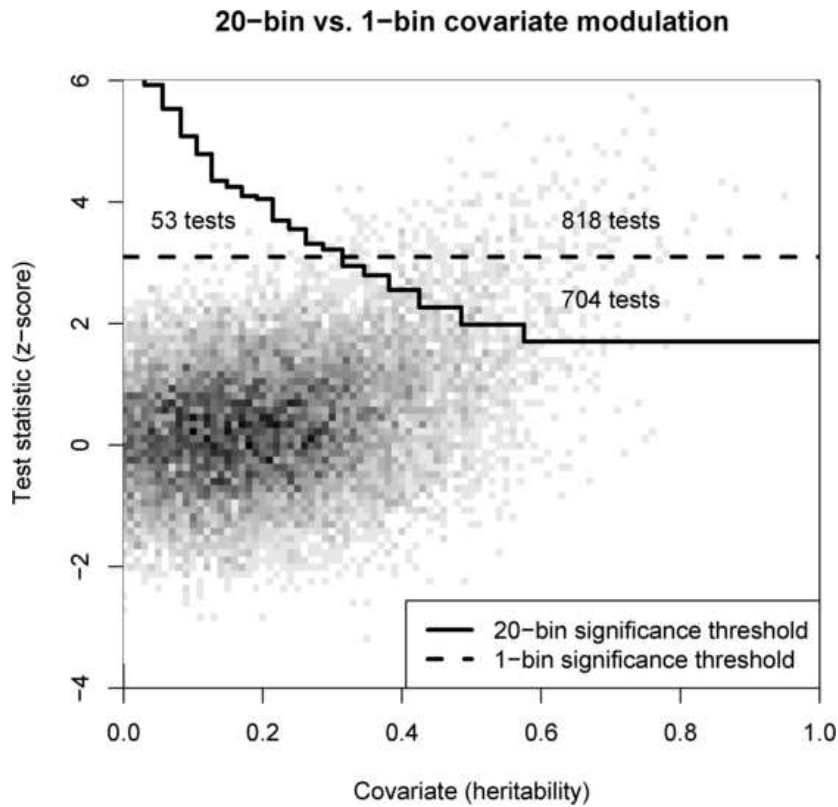

Fig. 2. *Two-dimensional histogram of pairs of covariate (heritability) and test statistic (z-score) for each of the 22317 tests (EST's). A darker gray tone indicates a higher density of points (tests) in each pixel. The solid step-like curve shows the significance threshold using the covariate-modulated, 20-bin posterior probability of $H_0$. The dashed line shows the significance threshold for the no-covariate, one-bin posterior probability. Both thresholds are at posterior probability level 0.05. 53 tests fall in the significance region of the one-bin method, but outside the significance region of the 20-bin method, 704 tests fall in the significance region of the 20-bin method, but outside the significance region of the one-bin method, while 818 tests are within the significance region of both methods. Using heritability has a very large impact.*

The one-bin model gives the posterior estimate $\hat{\pi}_0 = 0.701$, with approximate symmetric 95% credibility interval $(0.687, 0.714)$, for the overall proportion of true null hypotheses $\pi_0$. By comparison, the estimates $\hat{\pi}_0^s$ of Storey (2002) and $\hat{\pi}_0^c$ of Langaas et al. (2005) are $\hat{\pi}_0^s = 0.601$ and $\hat{\pi}_0^c = 0.615$, respectively, so it is possible that the one-bin uniform-beta mixture model gives a conservative estimate in this case.

As mentioned in Section 2, the ranking of genes according to the posterior probability of the alternative hypothesis may be different for the covariate-modulated posterior probability and for the one-bin, no-covariate posterior probability. This is illustrated in Figure 4, which shows the change in rankings for the top 100 genes for the 20-bin, covariate based method. Reshuffling is quite evident. Rank information is important for further analysis and experiments on the most promising EST's, with potential impact on drug-discovery plans.

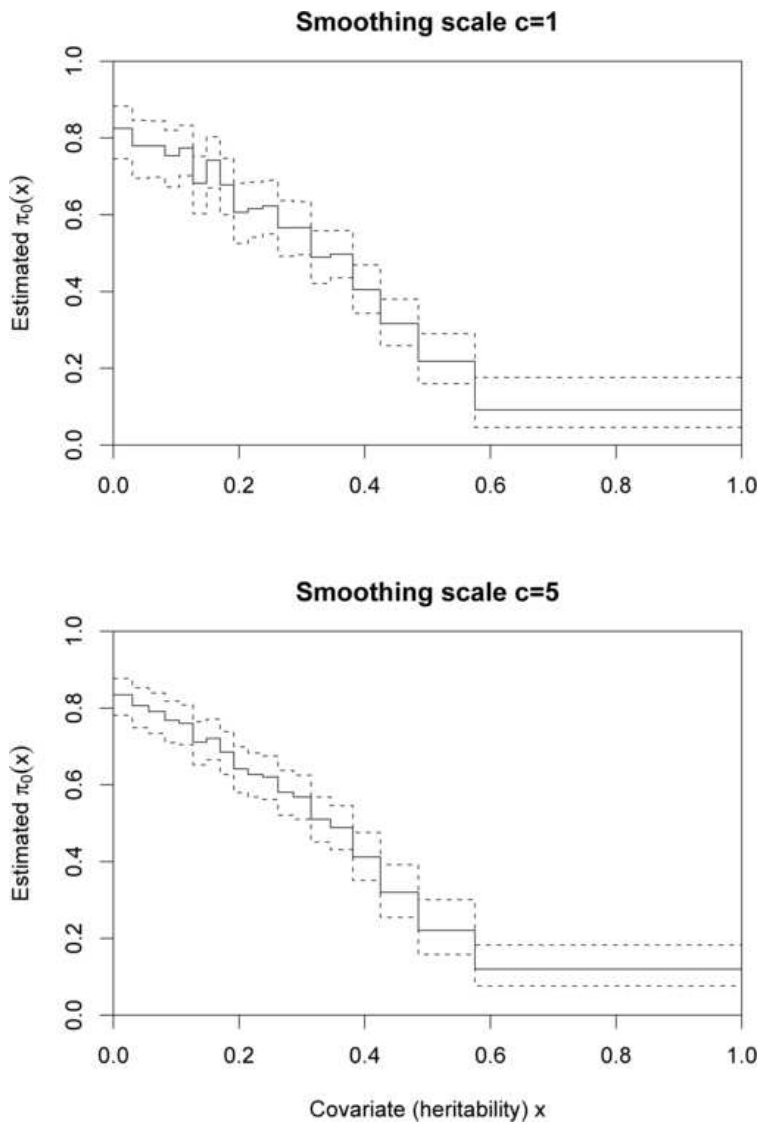

FIG. 3.   *Covariate-modulation functions for the eQTL data: Estimated $\pi_{0j}$ versus covariate (heritability) $x$. The steps correspond to bins. Two degrees of smoothing are shown (scaling factors $c=1$ and $c=5$, resp.). Pointwise 95% symmetric credibility intervals are shown as dashed lines.*

**5. Connections to other methods and optimality.** Several multiple testing methods use, for each test, the full likelihood of the data rather than a one-dimensional p-value/statistic summary. Use of the full likelihood implies that tests are reordered with respect to plain p-value ranking, thanks to information not conveyed by the corresponding univariate p-values. This happens also in our approach, but the additional information comes from external covariates, while in the full-likelihood (FL) methods there is a borrowing of strength between tests (internally).

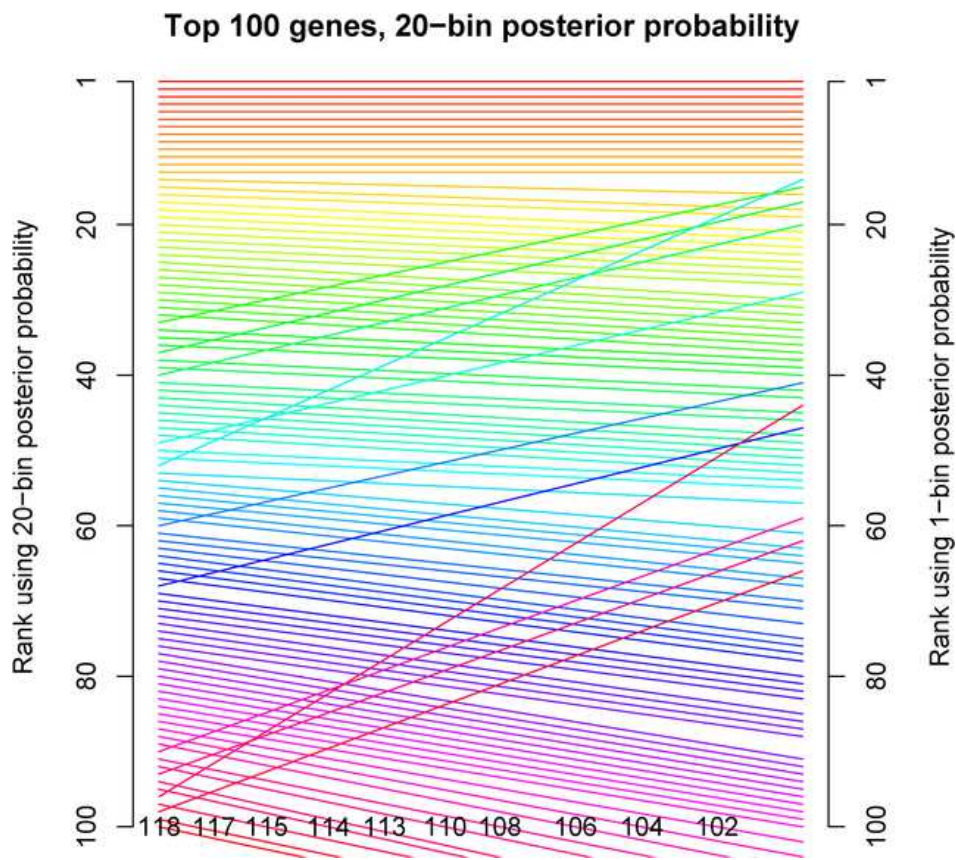

FIG. 4. *Comparison of ranks based on no-covariate (1-bin) posterior probability versus covariate-modulated (20-bin) posterior probability for the eQTL data. The EST's with lowest covariate-modulated posterior probability of $H_0$ are ranked along the left y-axis. Each segment ends on the right y-axis of the plot on the rank given to that EST by the one-bin method. Segments moving down indicate EST's which are more strongly cis-regulated according to the covariate-modulated posterior probability, while segments moving up indicate the opposite. Reshuffling of ranks is due to the merging of expression data with heritability knowledge. Some lines escape the plot; for these the final rank is written on the segment.*

In some cases our "external covariate" may be incorporated into the likelihood model (and thus become "internal" in our terminology), and the FL methods described below may be used directly. However, in many cases this will be impossible, impractical or unnatural. Our eQTL application below is an example where it would be inappropriate to use the covariate in this way, as was explained in Section 4.

Our method has the advantage that since everything is done conditionally on the covariate, no model for the covariate is needed. In this sense, our approach uses a minimal set of assumptions to provide a generally applicable way of using covariate information. FL methods are currently implemented for $k$-class comparisons (like the differential expression problem), while our methodology may readily be applied for any multiple testing problem where external covariate information is available.

Recently, Storey (2007) and Storey, Dai and Leek (2007) introduced a frequentist FL method called the Optimal Discovery Procedure (ODP). The ODP is a theoretically optimal procedure based on a generalized likelihood ratio thresholding function. The optimality goal is that of maximizing the expected number of true positives for each fixed expected number of false positives, where the former expectation is calculated under the subset where the alternative is true and the latter expectation is calculated under the subset where the null is true.

In the ODP method, two main steps are distinguished:

1. ordering tests by significance, and
2. choosing a cutoff to obtain a specified error rate.

The procedure is defined as follows. Suppose that we have $m$ significance tests performed on observed data sets $\mathbf{y}_1, \mathbf{y}_2, \ldots, \mathbf{y}_m$, where $\mathbf{y}_i$ is the vector of data for test $i$. Test $i$ has density $f_i$ under $H_0$ and density $g_i$ under $H_1$. Assume that $H_{0i}$ is true for $i = 1, \ldots, m_0$ and $H_{1i}$ is true for $i = m_0 + 1, \ldots, m$. The ODP for test $i$ is then given by the following significance thresholding function:

$$(6) \quad \mathcal{S}_{\text{ODP}}(\mathbf{y}_i) = \frac{f_1(\mathbf{y}_i) + \cdots + f_{m_0}(\mathbf{y}_i) + g_{m_0+1}(\mathbf{y}_i) + \cdots + g_m(\mathbf{y}_i)}{f_1(\mathbf{y}_i) + f_2(\mathbf{y}_i) + \cdots + f_{m_0}(\mathbf{y}_i)},$$

rejecting null hypothesis $H_{0i}$ if and only if $\mathcal{S}_{\text{ODP}}(\mathbf{y}_i) \geq \lambda$. Notice that step 1 above is here accomplished by the specification of the thresholding function (6), while step 2 is done by choosing $\lambda$. In our approach the cutoff is immediate to find, since we rank according to posterior probabilities.

Since the expression (6) for $\mathcal{S}_{\text{ODP}}(\mathbf{y}_i)$ involves several unknown quantities, the actual procedure is based on an estimated thresholding function $\hat{S}_{\text{ODP}}(\mathbf{y}_i)$. Storey, Dai and Leek (2007) propose to assume normality for the $\mathbf{y}_i$, and plug the maximum likelihood estimate into the Gaussian density functions to estimate $g_i$. [Note that $\hat{g}_{m_0+j}(y_i)$ is estimated using data $y_{m_0+j}$.] In addition, in the presence of nuisance parameters, it is necessary to provide a preliminary estimate of the summation of the true null densities. This may be done by ranking the tests by ordinary univariate significance criteria (e.g., p-values), estimating the proportion $\pi_0$ of true null hypotheses, and then deciding that the $m(1-\hat{\pi}_0)$ tests with smallest p-values have a true $H_1$, while the rest have a true $H_0$. Optimality is lost at this point, but satisfactory results are reported in practice. Choice of $\lambda$ is done using a bootstrapping procedure, described in the Online Supplement of Storey, Dai and Leek (2007).

The ODP procedure is closely related to our method. To see this, first disregard the presence of the external covariate, and assume that each null hypothesis $H_{0i}$ has the same probability $\pi_0$ of being true. In this case, it is shown in Storey (2007) (page 364) that the ODP thresholding function is equivalent to the thresholding function given by the posterior probability that the alternative hypothesis is true given the data $\mathbf{y}_i$. This implies that the ranking using this posterior probability is optimal in the ODP sense. Now, consider the use of external covariates. Our method is based on the posterior probability of $H_{1i}$ given a one-dimensional (p-value/test statistic) summary of the data *and* the covariate $x_i$. Thus, our method may be seen as a version of the ODP method in the setting where we only use a p-value/test statistic summary. Therefore, by optimality of the ODP, our method is optimal *within each bin* among methods based on one-dimensional summary statistics, but optimality over all bins is not guaranteed. It should be noted that *all* methods that are based on the posterior probability of the $H_{0i}$ ($H_{1i}$) can be seen as variants of the ODP method in this general sense.

It is a disadvantage of the ODP method to assume normality (or any other specific parametric model under $H_1$) of the measurements $\mathbf{y}$ and make an *ad hoc* preliminary guess of the status of each null hypotheses. In addition, the use of plug-in estimation is an issue. Our method is based on much fewer assumptions, basically only that valid p-values may be calculated, and that p-values/test statistics are independent of the covariate under the null. In this sense, our approach is unsupervised and data driven, and allows to incorporate covariates in a very simple and intuitive way.

Furthermore, our covariate-modulation ideas may be combined with the full-likelihood version of the ODP in at least two ways. One approach is to bin the data $\mathbf{y}_i$ by the covariate and estimate the ODP separately in each bin, extending a suggestion in Storey (2007) (page 353) to an external covariate. A problem with this approach is that is not clear how one could smooth between bins, which should be beneficial. Another approach is to first apply the ODP method to the data ignoring the covariate, calculate ODP-based adjusted p-values [as described in the Supplementary Material of Storey, Dai and Leek (2007), equation (10), page 5], and finally apply our method using the ODP-based p-values and the external covariate.

Other FL methods are based on hierarchical Bayesian mixture modeling. Relevant references include Baldi and Long (2001), Do, Müller and Tang (2005), Kendziorski et al. (2006), Lonnstedt and Speed (2002), Newton et al. (2004). As in our approach, inference is based on posterior probabilities of the null and alternative hypotheses. These methods require full Bayesian modeling and distributional assumptions, which are looked at with skepticism by practitioners used to unsupervised testing. As for the ODP, the hierarchical Bayesian mixture models may be combined with our approach. This can again be done by binning, and in the Bayesian approach smoothing between bins is now more immediately applicable. However, the use of our approach in a fully Bayesian setting may become computationally demanding.

The one-bin model differs from Efron's local FDR in that the one-bin model assumes a specific parametric mixture model for the p-values, whereas in the local FDR method the density of the p-values is estimated using parametric smoothing of the p-value histogram. A by-product of the estimation of the one-bin model is an estimate of $\pi_0$, the probability that a given null hypothesis is true (or, equivalently, the proportion of true null hypothesis). This is an interesting quantity in its own right, and several methods have been proposed for its estimation, such as the estimator of Storey (2002), and the convex decreasing estimator of Langaas, Lindqvist and Ferkingstad (2005). When estimating the full $B$-bin model, similarly an estimate of $\pi_{0j}$ can readily be found using the methods of Storey (2002) or Langaas, Lindqvist and Ferkingstad (2005) in each bin $j$, and this may be compared to the estimated $\pi_{0j}$ from the $B$-bin model.

**6. Simulation experiment.** This simulation experiment shows that the estimated covariate-modulated posterior probabilities of the null hypotheses are close to the true values in a controlled setting, and visualizes the effect of varying degrees of covariate modulation. Simulations were done for varying degrees of covariate-dependence and varying overall proportions of true null hypotheses.

The main procedure for the generation of each simulated data set is as follows: First, covariates $x_i$, $i = 1, \ldots, m$, are sampled iid from Unif$[0, 1]$, $m = 30000$. A covariate-modulation function $\pi_0(x)$ is specified (see below). For each $i$, we let $H_{0i}$ be true with probability $\pi_0(x_i)$, and let $H_{1i}$ be true with probability $1 - \pi_0(x_i)$. We sample $z_i$ from $N(0, 1)$ if $H_{0i}$ is true, and sample $z_i$ from $N(2,1)$ otherwise. The sampled $z_i$ are then test statistics for $m$ tests of $H_{0i} : \theta = 0$ vs. $H_{1i} : \theta = 2$, where $Z_i \sim N(\theta, 1)$.

As a first step in choosing a suitable covariate-modulation function, we decide on an overall level for the proportion of true null hypotheses, that is, we set $\bar{\pi}_0 \equiv \int_0^1 \pi_0(x)\,dx$ to some fixed number. The strength of the dependence on the covariate is determined by the steepness of the function $\pi_0(x)$. We have chosen the following parametric form for $\pi_0(x)$:

$$(7) \qquad \pi_0(x) = \exp(-\alpha - (\beta - \alpha)x^\gamma),$$

where $\alpha$, $\beta$ and $\gamma$ are positive constants, and $\pi_0(x)$ is decreasing if and only if $\alpha > \beta$. We have confined our simulation experiment to decreasing $\pi_0(x)$.

For each value of $\bar{\pi}_0$, we set the strength of the covariate-dependence by choosing appropriate $\alpha$ and $\beta$, implicitly setting $\pi_0(0)$ and $\pi_0(1)$ since $\alpha = -\log(\pi_0(0))$ and $\beta = -\log(\pi_0(1))$, and then choosing $\gamma$ such that $\int_0^1 \exp(-\alpha - (\beta - \alpha)x^\gamma)\,dx = \bar{\pi}_0$.

The simulation results reported below are all from 1000 runs of the algorithm described in Section 3. Pointwise medians, 5% quantiles and 95% quantiles were calculated based on the 1000 runs. In each run a total number of 30000 test statistics were simulated. In each case, one-bin no-covariate and 10-bin covariate-modulated posterior probabilities were calculated for the same data sets. True posterior probabilities were calculated as described in the Supplementary Material [Ferkingstad, Frigessi, Rue, Thorleifsson and Kong (2008)].

We first consider two simulated data sets with relatively weak covariate-modulation. We choose $\alpha$, $\beta$ and $\gamma$ in (7) such that $\bar{\pi}_0 = 0.5$, $\pi_0(0) = 0.55$ and $\pi_0(1) = 0.45$. The top row in Figure 5 show the estimated posterior probabilities for the 10-bin and one-bin methods, compared to the true posterior probabilities, in bins 1, 3, 5, 7 and 9. The red curves show the median estimated covariate-modulated posterior probabilities (solid lines) together with 0.05 and 0.95 quantiles (dashed lines) from the 1000 runs. The green curves show the true posterior probabilities. Only posterior probabilities for p-values in the range $[0, 0.1]$ are shown, since the posterior probabilities are too high to be of interest outside this range. Notice first that the one-bin posterior probability is invariant over different bins, since this does not depend on covariate-information. However, the estimated covariate-modulated posterior probability and the true posterior probability vary over different bins. From the two top rows of Figure 5, we see that the estimated 10-bin posterior probability is very close to the truth. The one-bin estimated posterior probability is a good estimate for bins corresponding to a moderate value of the covariate (bin 5), but deteriorates for more extreme values of the covariate (other bins). This is as expected: Since covariate modulation is quite weak, the information in the covariate should not change the discoveries. This result is consistent with what we saw for a second real data set, as reported in the Supplementary Material [Ferkingstad, Frigessi, Rue, Thorleifsson and Kong (2008)].

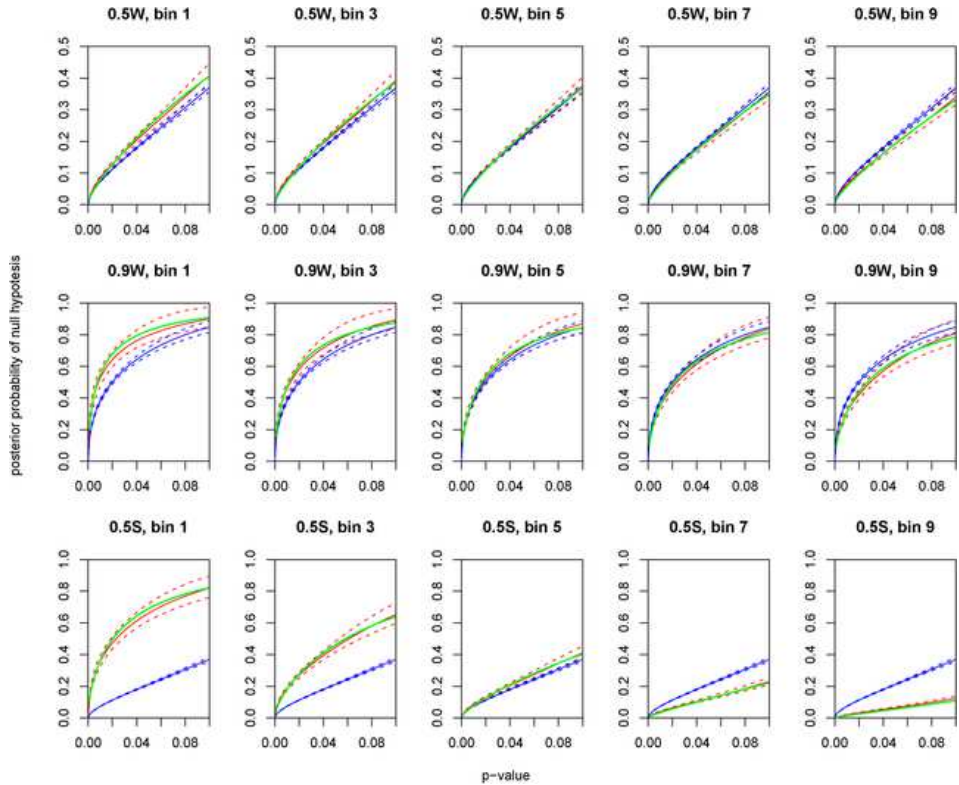

Fig. 5. *Simulated data sets. The top row shows estimated posterior probabilities for $\bar{\pi}_0 = 0.5$ with weak covariate-modulation, the middle row shows results for $\bar{\pi}_0 = 0.9$ with weak covariate-modulation and the bottom row shows results for $\bar{\pi}_0 = 0.5$ with strong covariate-modulation. Ten bins were used in the calculations, but only bins 1, 3, 5, 7 and 9 are shown here. For each bin, the bold green curve shows the true posterior probabilities. The red curves show estimated 10-bin covariate-modulated posterior probabilities, where the solid curve is the median of the 1000 simulations, and the dashed curves are 0.05 and 0.95 quantiles. Similarly, the blue curves show estimated no-covariate (one-bin) posterior probabilities (median with 0.05 and 0.95 quantiles).*

As the second case of weak covariate-modulation, we choose $\bar{\pi}_0 = 0.9$, $\pi_0(0) = 0.95$ and $\pi_0(1) = 0.85$. Results, shown in the middle row of Figure 5, are similar to what we saw in the first case. Again, it is seen that the covariate-based estimation method gives a very close approximation to the truth, while the noncovariate-based method misses off slightly, particularly for the bins corresponding to extreme (particularly small or large) values of the covariate.

Finally, we consider a simulated data set where there is strong covariate-modulation. Here, we have chosen $\bar{\pi}_0 = 0.5$, $\pi_0(0) = 0.9$ and $\pi_0(1) = 0.1$. The results are shown in the bottom row of Figure 5. We again see that the estimated 10-bin covariate-modulated posterior probabilities are very close to the true values. Clearly, the covariate-modulation method does well in capturing the information in the covariate in this case. There is a big difference between the 10-bin posterior probability and the no-covariate method. The no-covariate method does not work well in any bins (except possibly bin 5). When covariate information is available, using it, even in an unsupervised fashion, has a potentially important impact.

**7. Discussion.** We defined the covariate-modulated posterior probability as $P(H_{0i}|p_i, x_i)$, where $p_i$ is the $i$th p-value and $x_i$ the corresponding covariate for $i = 1, \ldots, m$ hypothesis tests. More generally, we can consider $P(H_{0i}|p_i, \mathbf{x})$, where $\mathbf{x} = (x_1, x_2, \ldots, x_m)$ is a vector of covariates.

We use the assumption that $g(z_i|H_{0i}, x_i) = g(z_i|H_{0i})$, where $g$ denotes the density of the test statistics. However, the method could also be adapted to the case where $g(z|H_{0i}, x_i)$ is *known* as a function of $x_i$.

Furthermore, additional beta components can be added to the mixture (3). Also, it is possible to incorporate measurement error in the covariate, by adding a further level to the Bayesian hierarchical model.

In the Supplementary Material [Ferkingstad, Frigessi, Rue, Thorleifsson and Kong (2008)] we test for differentially expressed genes, between breast cancer tumors with and without a mutation in the gene TP53. Here, we use genome-wide CGH copy number alteration as the covariate modulating the a priori belief in each null hypothesis. This data set is chosen to illustrate a case where the information contained in the covariate is limited, but its use does not corrupt the results.

An important application where a covariate can play an important role is genome-wide association studies where hundreds of thousands of SNPs (single nucleotide polymorphisms) are tested for association to quantitative or qualitative traits. With gene expression traits, p-values calculated for SNPs residing in the neighborhood of the corresponding genes (cis-variants) could again be evaluated using our approach. Here both the heritability and the cis linkage score could be used as covariates, either individually or jointly. A related application is an association study for a disease trait (e.g., diabetes). Here the notion of a cis variant does not apply and it is expected that a large number of SNPs, but nonetheless a very small fraction of all the SNPs typed, could be associated to the disease. A common design is a case-control study, and the association scores for the hundreds of thousands of SNPs genotyped and tested have to be evaluated taking multiple comparisons into account. Indeed, it has been proposed that the linkage scores resulting from a family study of the disease could be used to assist the analysis through differential weighting of the p-values [Roeder et al. (2006), Genovese et al. (2006)] with the weights working like prior probabilities. The method we proposed here is an alternative to the weighting method. This application also raises the possibility that instead of modeling the distribution of the p-values under $H_1$ directly, one can model the effect sizes of the association; given a distribution of the effect sizes, there is a corresponding distribution for the p-values. Modeling this way can in theory provide for each SNP, at the end of the analysis, not only the probability that the alternative hypothesis is true, but also the posterior distribution of its effect size. The shrinkage characteristics of such methods could be a potential solution to the winner's curse, where the observed effect sizes for the SNPs that showed the strongest associations tend to be biased upward even when they are true positives.

**Acknowledgments.** We thank the Editor, an Associate Editor and an anonymous referee for helpful comments that have led to an improved paper.

## SUPPLEMENTARY MATERIAL

**Unsupervised empirical Bayesian multiple testing with external covariates** (doi: 10.1214/08-AOAS158SUPP; .pdf).

E. Ferkingstad
Department of Biostatistics
University of Oslo
PO box 1122 Blindern
N-0317 Oslo
Norway
E-mail: egil.ferkingstad@medisin.uio.no

A. Frigessi
(sfi)$^2$—Statistics for Innovation
The Norwegian Computing Centre
PO box 114 Blindern
N-0314 Oslo
Norway
E-mail: arnoldo.frigessi@medisin.uio.no

H. Rue
Department of Mathematical Sciences
Norwegian University of Science and Technology
N-7194 Trondheim
Norway
E-mail: havard.rue@math.ntnu.no

G. Thorleifsson
A. Kong
Decode Genetics
Sturlugata 8
IS-101 Reykjavik
Iceland
E-mail: gudmar.thorleifsson@decode.is
     augustine.kong@decode.is